\documentclass[12pt]{article}
\usepackage{epsfig}
\usepackage{cite}

\begin{document}
\begin{titlepage}

\centerline{\large\bf Center vortex model for the infrared sector of SU(3)}
\vspace{0.2cm}
\centerline{\large\bf Yang-Mills theory: Topological susceptibility}

\bigskip
\centerline{M.~Engelhardt\footnote{\tt email:\ engel@nmsu.edu} }
\vspace{0.2 true cm}
\centerline{\em Department of Physics, New Mexico State University}
\centerline{\em Las Cruces, NM 88003, USA}

\abstract{The topological susceptibility of the $SU(3)$ random vortex
world-surface ensemble, an effective model of infrared Yang-Mills dynamics,
is investigated. The model is implemented by composing vortex world-surfaces
of elementary squares on a hypercubic lattice, supplemented by an
appropriate specification of vortex color structure on the world-surfaces.
Topological charge is generated in this picture by writhe and
self-intersection of the vortex world-surfaces. Systematic uncertainties
in the evaluation of the topological charge, engendered by the hypercubic
construction, are discussed. Results for the topological susceptibility
are reported as a function of temperature and compared to corresponding
measurements in $SU(3)$ lattice Yang-Mills theory. In the confined phase,
the topological susceptibility of the random vortex world-surface
ensemble appears quantitatively consistent with Yang-Mills theory.
As the temperature is raised into the deconfined regime, the topological
susceptibility falls off rapidly, but significantly less so than in $SU(3)$
lattice Yang-Mills theory. Possible causes of this deviation, ranging
from artefacts of the hypercubic description to more physical sources,
such as the adopted vortex dynamics, are discussed.}

\vspace{1cm}

{\footnotesize PACS: 12.38.Aw, 12.38.Mh, 12.40.-y}

{\footnotesize Keywords: Center vortices, infrared effective theory,
confinement, topology}

\end{titlepage}

\section{Introduction}
A model of the QCD vacuum which has been successful in capturing the
fundamental phenomena characterizing the strong interaction in its
infrared, strongly coupled sector is the center vortex picture. It
assumes that the long-wavelength modes of the gluon field are collectively
organized into randomly distributed, percolating tubes of quantized
chromodynamic flux in three-dimensional space; early in its development
\cite{hooft,aharonov,cornold,map1,mack,map2,mapi,niol,aol1,aol2,ole,tomb},
this picture was often referred to, employing vivid imagery, as the
``spaghetti vacuum''. The aforementioned tubes of flux are termed
``center vortices'' since flux quantization is determined by the center
of the underlying gauge group (detailed definitions are given in
section~\ref{vordef}). Contrary to spaghetti, vortices have no open
ends; this is an expression of the Bianchi identity, i.e., continuity
of flux (modulo Abelian magnetic monopoles). Also contrary to
spaghetti, vortices can move through one another, i.e., their
world-surfaces can intersect. The vortex picture, including its
relation to other models of the QCD vacuum, has been reviewed in
\cite{jg3,latrev}.

While the vortex picture was originally conceived specifically as a possible
mechanism of quark confinement, more recent developments have shown that it
also provides viable explanations of the other two central phenomena observed
in the low-energy sector of the strong interaction, namely, the spontaneous
breaking of chiral symmetry and the axial $U_A (1)$ anomaly (the latter
representing the focus of the study presented here). The vortex picture thus
provides a comprehensive, consistent account of the gross features of the
strong interaction vacuum. Two main lines of investigation have contributed
to these developments. On the one hand, sparked by the inception of
practicable algorithms for the detection of vortices in lattice Yang-Mills
configurations \cite{jg1,jg2,df2}, lattice studies of the vortex content of
Yang-Mills theory and the effects it induces were carried out
\cite{jg1,jg2,df2,per,df1,rb,hellfabolej,su2corr,su3corr}. On the other hand,
an infrared effective model based directly on center vortex degrees of
freedom with a simplified effective dynamics was introduced to complement the
lattice studies and expand on the range of vortex physics that could be
accessed quantitatively \cite{m1,m2,m3,su3conf,su3bary,su3freee,su4,sp2}.

In the lattice Yang-Mills approach, identifying center vortices within
lattice configurations containing the full Yang-Mills dynamics is a complex
pattern recognition problem. While center vortices are, in principle, defined
gauge-invariantly via their effect on Wilson loops, cf.~section~\ref{vordef},
this pattern recognition problem is usually handled by adopting particular
gauges which facilitate projecting out the vortex content of a given
configuration. Two classes of gauges which have been employed in this
respect are maximal center gauges \cite{jg1,jg2} and Laplacian center
gauges \cite{df2}. On the basis of these methods, lattice Yang-Mills
studies have demonstrated center dominance, i.e., that the vortex content
of lattice Yang-Mills configurations fully accounts for the asymptotic
string tension, both at zero temperature \cite{jg1,jg2,df2} and at finite
temperatures \cite{per}; the deconfining phase transition is revealed as
a percolation transition (in certain three-dimensional slices of space-time)
in the vortex picture \cite{per}. Moreover, vortices account for the
topological content of the Yang-Mills ensemble \cite{df1,rb}. The study of
the chiral symmetry breaking effects induced by vortices via the low-lying
modes of the Dirac operator has proven to be technically more challenging
due to the fact that center vortex configurations projected from full
lattice Yang-Mills configurations are not smooth; nevertheless, it has been
shown that chiral symmetry breaking disappears (along with topological
charge and confinement) when vortices are removed from the full lattice
configurations \cite{df1,df2}, and a detailed study employing
asqtad quarks \cite{hellfabolej} has demonstrated the emergence
of a dense low-lying Dirac eigenvalue spectrum in the ensemble of
vortex configurations projected from their full Yang-Mills counterparts.

These findings have been complemented by further recent investigations
focusing on correlations between center vortices and low-lying overlap Dirac
operator eigenmodes \cite{su2corr,su3corr,hellfabolej}, which, on the other
hand, can be tied to the topological charge density. Of related interest are
studies of the connection between center vortices and other topological
charge carriers arising in Yang-Mills theory; a detailed analysis of
the vortex content of calorons was presented recently in \cite{zhang}.
Finally, a more formal issue which arises where correlations between
center vortices and Dirac operator eigenmodes are concerned is the
form of the index theorem in the presence of vortices; the non-smoothness
of vortex gauge fields already alluded to further above may introduce
complications in this respect. This has been investigated recently
in \cite{hellfab1,hellfab2}. Some observations on the related question
of the quantization properties of global topological charge are made below
in section~\ref{quantsec}.

As already indicated, the lattice Yang-Mills studies of vortex physics
highlighted above have been complemented by the formulation of a
corresponding infrared effective model of center vortices. Since
vortices represent lines of chromodynamic flux in three space
dimensions, they correspondingly are described by two-dimensional
world-surfaces in four-dimensional space-time. Implementing the notion
that center vortices are randomly distributed, a random vortex
world-surface model in Euclidean space-time was introduced and studied in
\cite{m1,m2,m3,su3conf,su3bary,su3freee,su4,sp2}. Concentrating initially
on an underlying $SU(2)$ gauge group, the confinement properties, including
the finite temperature phase transition to a deconfined phase \cite{m1},
the topological susceptibility \cite{m2} and the (quenched) chiral
condensate \cite{m3} were found to quantitatively reproduce the
corresponding features in $SU(2)$ lattice Yang-Mills theory.
Subsequently, the model was generalized to other gauge groups, the
confinement properties being investigated not only for the $SU(3)$
case \cite{su3conf,su3bary,su3freee}, but also for $SU(4)$ \cite{su4}
and $Sp(2)$ \cite{sp2}. For the most relevant case of an underlying
$SU(3)$ gauge group, a weakly first order deconfinement transition
\cite{su3conf,su3freee} and a Y-law for the baryonic static potential
\cite{su3bary} were found, in accordance with $SU(3)$ lattice Yang-Mills
theory. The present work continues the investigation of the $SU(3)$
model, focusing on the topological susceptibility, which is instrumental
in determining, via the axial $U_A (1)$ anomaly, the mass of the
$\eta^{\prime } $ meson. Preliminary accounts of this work have been
given in \cite{mla08,mco8}.

\section{Modeling center vortices}

\subsection{Center vortex degrees of freedom}
\label{vordef}
The vortex picture of the strong interaction vacuum assumes that the
relevant infrared gluonic degrees of freedom are center vortices. On
infrared length scales, center vortices are closed lines of quantized
chromomagnetic flux in three space dimensions. They are therefore
described by closed two-dimensional world-surfaces in four-dimensional
space-time. Their flux is quantized according to the center of the gauge
group; if one evaluates a Wilson loop $W$ encircling an $SU(3)$ vortex flux,
one obtains one of the nontrivial\footnote{Of course, the trivial unit
center element signals that no flux is present.} center elements of the
$SU(3)$ group, i.e.,
\begin{equation}
W=\exp (\pm 2\pi i/3) \ .
\label{centel}
\end{equation}
Note that the two center elements in question are complex conjugates of
one another, implying that the $SU(3)$ gauge group only really allows
for one type of vortex flux, the two possible space-time orientations
of which determine which center element is measured.

Note furthermore that the specific structure of the $SU(3)$ center also
allows for vortex branching. A vortex flux associated with
$W=\exp (2\pi i/3)$ branching into two vortex fluxes each associated with
$W=\exp (-2\pi i/3)$ is compatible with the Bianchi constraint, i.e., flux
continuity modulo Abelian magnetic monopoles; evaluating a Wilson loop
encircling the latter two fluxes yields
$W=\exp (-2\pi i/3) \cdot \exp (-2\pi i/3)=\exp (2\pi i/3)$,
just as for the original flux.

Viewing center vortices as infrared effective degrees of freedom implies
that their space-time location is only determined to an accuracy limited
by the ultraviolet cutoff. Equivalently, if one sufficiently increases
the space-time resolution, it is appropriate to represent center vortices
as thickened tubes in three space dimensions, or correspondingly thickened
world-surfaces in space-time. This thickness, encoding the ultraviolet
cutoff, is relevant for medium-range phenomena such as Casimir scaling of
the static quark potential at intermediate distances \cite{cassc,greentok}.
It plays a role in the construction of an infrared effective vortex
dynamics, cf.~section~\ref{mcdyn}.

\subsection{Vortex field strength}
To evaluate the topological charge
\begin{equation}
Q=\frac{1}{32\pi^{2} } \int d^4 x \, \epsilon_{\mu \nu \lambda \tau } \
\mbox{Tr} \ F_{\mu \nu } F_{\lambda \tau }
\end{equation}
of center vortices, it is necessary to associate a chromodynamic field
strength tensor $F_{\mu \nu } $ with them. While it will not be necessary
to give a general construction of $F_{\mu \nu } $ for an arbitrary vortex
configuration \cite{contvort}, a few of its properties need to be specified
for the developments further below.

A vortex world-surface element running in the $\rho $ and $\sigma $
directions carries a field strength $F_{\mu \nu } $, localized on the
world-surface, such that the $\mu $ and $\nu $ directions are perpendicular
to the $\rho $ and $\sigma $ directions \cite{contvort,m2}. Apart from this
gauge-invariant statement, the field strength also has a direction in color
space, which can be rotated by gauge transformations. It is convenient to
cast vortex configurations in an Abelian gauge, i.e., the $3\times 3$ color
matrix $F_{\mu \nu } $ will be chosen diagonal. Vortex color structure can
be usefully characterized by eliminating the space-time details of
the vortex field strength and considering only the color direction
$T(r)$ of the vortex at the position $r$,
\begin{equation}
\frac{2\pi }{3} T(r) = \frac{1}{2} \int_{S_r } F_{\mu \nu } \,
d^2 S_{\mu \nu } = \int_{\partial S_r } A_{\mu } \, dx_{\mu } \ ,
\end{equation}
where $S_r $ is a two-dimensional surface intersecting the vortex at $r$
(but intersecting no other fluxes) and $A_{\mu } $ is a (diagonal) gauge
field generating the field strength $F_{\mu \nu } $. In terms of a
parametrization $x(s_1 ,s_2 )$ of $S_r $, the oriented surface element
is given by $d^2 S_{\mu \nu } =\epsilon_{\kappa \lambda }
(\partial x_{\mu } /\partial s_{\kappa } )
(\partial x_{\nu } /\partial s_{\lambda } ) ds_1 ds_2 $.
In terms of $T$, the Wilson loop along the contour
$\partial S_r $ encircling the vortex is simply given by
\begin{equation}
W=\frac{1}{3} \, \mbox{Tr} \, \exp (2\pi i T/3) \ ,
\end{equation}
i.e., $T$ has integer entries. As one travels along the vortex
world-surface, $T$ remains constant except possibly at lines on the
surface at which $T$ switches from one color direction to another,
in a manner which must of course be compatible with the Bianchi
constraint. In fact, such switches in the color direction $T$ are
unavoidable on generic vortex world-surfaces. In the simpler case of
an $SU(2)$ gauge group \cite{m2}, this comes about purely due to the
nonorientability of the surfaces. Nonorientability implies that there
must be lines on the surfaces at which the orientation of vortex flux
is inverted and $T$ therefore displays a discontinuity. In the $SU(3)$
case considered here, the picture is complicated by the branched nature
of the surfaces, cf.~further below. Moreover, while in the $SU(2)$ case,
the choice of the set of allowed color directions $T$ is essentially unique,
in the $SU(3)$ case, one has a certain amount of (gauge) freedom in the
choice of the set of allowed color directions. These options, leading to
different patterns of discontinuities on the vortex world-surfaces,
will be discussed further in the next section.

In more physical terms, a discontinuity in the color direction of vortex
flux described by $T$ implies the presence of a source or sink of that
flux, i.e., the presence of an Abelian magnetic monopole world-line
on the vortex world-surface. In view of the fact that such discontinuities
in general cannot be avoided, Abelian magnetic monopoles represent
intrinsic features of vortex configurations cast in Abelian gauges.
This is, of course, the character of Abelian gauges; rotations of the
field strength tensor in color space, which in general occur continuously
as a function of space-time location, are compressed into singular jumps.
The precise locations of the monopole world-lines on the vortex
world-surfaces can be shifted by gauge transformations, but certain
topological characteristics of these singularities are gauge-invariant
(they are, e.g., in general non-contractible), and influence, in particular,
the topological charge\footnote{Indeed, on a torus with periodic boundary
conditions (nontrivial boundary conditions such as torus twist require
additional consideration), globally oriented vortex world-surfaces, i.e.,
ones devoid of monopoles, carry no global topological charge
\cite{contvort}.}.

\subsection{Vortex color structure}
\label{colstruc}
\begin{figure}
\vspace{-0.1cm}
\centerline{
\epsfig{file=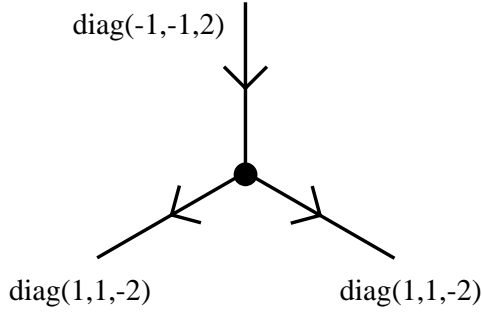,width=8cm}
}
\caption{Vortex color structure at branchings for a minimal choice of
the set of allowed color generators $T$, cf.~(\ref{minchoice}).
The restricted set of available $T$ forces a monopole to appear at
the branching; the difference between incoming and outgoing fluxes
is $\mbox{diag} (-1,-1,2) - 2\cdot \mbox{diag} (1,1,-2) =
\mbox{diag} (-3,-3,6) = \mbox{diag} (-3,0,3) + \mbox{diag} (0,-3,3)$,
i.e., monopole flux which can be further decomposed into two
elementary monopoles in two separate $SU(2)$ subgroups of $SU(3)$,
as indicated by the second equality.}
\label{branchminfig}
\end{figure}

As indicated above, for the $SU(3)$ gauge group, one has different options
in the choice of the set of allowed color directions $T$ on vortex
world-surfaces, corresponding to a residual freedom in the choice of
Abelian gauge. Consider, to begin with, a minimal set, i.e., let
\begin{equation}
T \in \left\{ \pm \, \mbox{diag} (1,1,-2) \right\} \ .
\label{minchoice}
\end{equation}
This is sufficient to generate both nontrivial center elements of the
$SU(3)$ group, cf.~(\ref{centel}), i.e., both possible orientations of
vortex flux. Consider now the occurrence of monopoles. Contrary to the
$SU(2)$ case \cite{m2}, in this description, monopoles cannot occur
away from branchings, since the flux required to switch from
$T=\mbox{diag} (1,1,-2)$ to $T=\mbox{diag} (-1,-1,2)$ does not
correspond to a possible Abelian magnetic monopole flux (which would
be described by diagonal elements which are integer multiples of $3$ in
the convention used here). On the other hand, monopoles {\em must} occur
at branchings, cf.~Fig.~\ref{branchminfig}.

\begin{figure}
\vspace{-0.1cm}
\centerline{
\epsfig{file=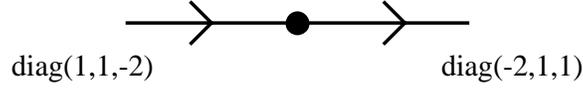,width=8cm}
\vspace{-0.5cm}
}
\caption{Possible monopole not associated with branching for a non-minimal
choice of the set of allowed color generators $T$, cf.~(\ref{nonmin}).
The difference between incoming and outgoing fluxes is
$\mbox{diag} (1,1,-2) - \mbox{diag} (-2,1,1) =  \mbox{diag} (3,0,-3)$,
i.e., a monopole flux.}
\label{flipnonminfig}
\end{figure}

\begin{figure}
\vspace{-0.1cm}
\centerline{
\epsfig{file=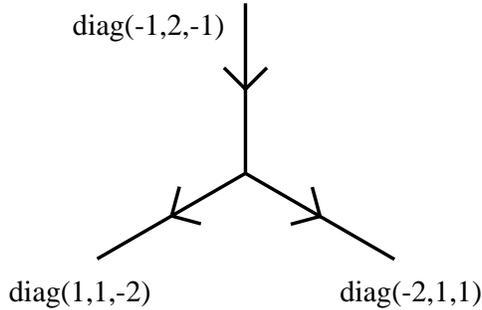,width=8cm}
}
\caption{Possible vortex color structure at branchings for a non-minimal
choice of the set of allowed color generators $T$, cf.~(\ref{nonmin}).
One can choose incoming and outgoing fluxes such that their difference
vanishes, $\mbox{diag} (-1,2,-1) - \mbox{diag} (1,1,-2) -
\mbox{diag} (-2,1,1) = 0$. In this description, one thus has the
freedom to disassociate monopoles from branchings.}
\label{branchnonminfig}
\end{figure}

Consider, on the other hand, a non-minimal, more symmetric choice,
\begin{equation}
T \in \left\{ \pm \, \mbox{diag} (1,1,-2), \pm \, \mbox{diag} (1,-2,1),
\pm \, \mbox{diag} (-2,1,1) \right\} \ .
\label{nonmin}
\end{equation}
This description, introduced in \cite{cw1,cw2,cw3}, allows for more
flexibility; monopoles can occur away from branchings,
cf.~Fig.~\ref{flipnonminfig}, and branchings are not necessarily
associated with monopoles, cf.~Fig.~\ref{branchnonminfig}. In fact,
this description affords the possibility of deforming monopole world-lines
such that they never coincide with vortex world-surface branching lines,
except for, at most, isolated crossings of the former and the latter. This
property singles out the choice (\ref{nonmin}) as the one best suited for
the purpose of evaluating the topological charge of $SU(3)$ vortex
configurations, cf.~section~\ref{ambisec}.

\subsection{Vortex world-surfaces on a hypercubic lattice}
\label{lattsurf}
In order to arrive at a practical scheme of generating model vortex
world-surface ensembles, cf.~section~\ref{mcdyn}, the world-surfaces will
be composed of elementary squares on a hypercubic lattice. One can
then describe a vortex configuration by recording the chromodynamic
flux associated with each elementary square in the lattice. Associate
the lattice elementary square extending from the lattice site $x$ into
the positive $\mu $ and $\nu $ directions with a $3\times 3$ color
matrix $p_{\mu \nu } (x)$, where either $p_{\mu \nu } (x) =0$ 
(indicating the absence of vortex flux) or $p_{\mu \nu } (x) =T$
with $T$ indicating the color orientation of the vortex flux on
the square in question, as introduced in the previous section,
cf.~(\ref{nonmin}). Note that the order of indices defines a sense of curl
and, accordingly, $p_{\nu \mu } (x)$ and $p_{\mu \nu } (x)$ are related by
space-time inversion, i.e., $p_{\nu \mu } (x) = -p_{\mu \nu } (x)$.
In practice, it is thus sufficient to record $p_{\mu \nu } (x)$ for
$\mu < \nu $.

An operation repeatedly used in the following is an {\em elementary
cube transformation}, which locally deforms a given vortex world-surface
configuration on the lattice into a new configuration as follows. An
additional closed vortex flux, of the shape of the surface of an
elementary three-dimensional cube (the smallest possible closed
vortex world-surface), and associated with one of the possible color
orientations $T$ from (\ref{nonmin}), is superimposed onto the original
configuration. This creates a new configuration while preserving the
Bianchi constraint, i.e., continuity of flux modulo Abelian magnetic
monopoles. Specifically, if the elementary cube in question extends from
the lattice site $x$ into the positive $\mu $, $\nu $ and $\lambda $
directions, the transformation effects
\begin{equation}
\begin{array}{ll}
p_{\mu\nu}(x) \rightarrow \mbox{Mod} \, (p_{\mu\nu}(x) + T)
\, , \ \ \ \ &
p_{\mu\nu}(x+e_\lambda) \rightarrow \mbox{Mod} \,
(p_{\mu\nu}(x + e_\lambda) - T) \\
p_{\nu \lambda}(x) \rightarrow \mbox{Mod} \, (p_{\nu \lambda}(x) + T)
\, , \ \ \ \ &
p_{\nu \lambda}(x+e_\mu) \rightarrow \mbox{Mod} \,
(p_{\nu \lambda}(x+e_\mu) - T) \\
p_{\lambda\mu}(x) \rightarrow \mbox{Mod} \, (p_{\lambda\mu}(x) + T)
\, , \ \ \ \ &
p_{\lambda\mu}(x+e_\nu) \rightarrow \mbox{Mod} \,
(p_{\lambda\mu}(x+e_\nu) - T)
\end{array}
\end{equation}
where $\mbox{Mod} $ denotes a generalized modulo operation, acting on
diagonal $3\times 3$ color matrices, which maps its argument back into
the set of allowed color orientations, cf.~(\ref{nonmin}); it acts as
follows:
\begin{equation}
\begin{array}{lcl}
\mbox{Mod} (P) = 0 & \ \ \ \mbox{if} \ \ \ & \det P =0 \nonumber \\
\mbox{Mod} (P) = -P/2 & \ \ \ \mbox{if} \ \ \ & |\det P| =16 \\
\mbox{Mod} (P) = P & \mbox{else} & \nonumber
\end{array}
\end{equation}
Note that the first two alternatives in general induce Abelian magnetic
monopole lines in the transformed configuration.

Elementary cube transformations will be employed further below both
as Monte Carlo updates\footnote{Strictly speaking, purely as a matter
of technical convenience, Monte Carlo updates will act directly on the
reduced quantities $q_{\mu \nu } (x)$ introduced in (\ref{qred}), which
carry only part of the information contained in $p_{\mu \nu } (x)$; the
action is entirely analogous and can be unambiguously inferred from the
definition given here.} in the generation of vortex world-surface
ensembles, as well as in the process of measuring the topological charge.

\section{Vortex topological charge}
\subsection{Origin of vortex topological charge density}
The topological charge of vortex world-surfaces results from world-surface
intersections and world-surface writhe \cite{m2,cw1,cw4,contvort,bruck}. If
one considers idealized, infinitely thin surfaces in four-dimensional
space-time, intersections occur at isolated points in space-time, whereas
writhe in general is continuously distributed along surfaces. An illustrative
example is given in \cite{bruck}. In an infrared effective framework,
where locations cannot be specified to higher accuracy than given by
the ultraviolet cutoff, these features are smeared out over the
corresponding length scale (which in practice can be identified with
the vortex thickness).

On the other hand, if for modeling purposes one composes vortex
world-surfaces from elementary squares on a hypercubic lattice, as
will be done below, additional considerations must be taken into account.
In such a setting, topological charge can only be generated at lattice
sites, since these are the only space-time points at which surface
elements can meet such that their tangent vectors span all four
space-time dimensions\footnote{Recall that a vortex world-surface
running in the $\rho $ and $\sigma $ directions is associated with a
field strength $F_{\mu \nu } $ such that the $\mu $ and $\nu $ directions
are perpendicular to the $\rho $ and $\sigma $ directions. Therefore,
to generate a nonvanishing topological density
$\epsilon_{\mu \nu \lambda \tau } \ \mbox{Tr} \
F_{\mu \nu } F_{\lambda \tau } $, surface elements must meet such
that their tangent vectors span all four space-time dimensions.}.
Thus, also vortex writhe becomes concentrated on space-time points
instead of being continuously distributed on vortex world-surfaces.

At first sight, it would therefore seem that the topological charge $Q$
of hypercubic model surfaces can be evaluated simply by considering all
lattice sites $x$, and at each site counting pairs of mutually orthogonal
elementary squares meeting there, appropriately weighted by the
associated chromodynamic flux,
\begin{equation}
Q=\sum_{x} q(x) \ , \ \ \ \ \ \ \ 
q(x) = \frac{1}{288} \sum_{\mu < \nu } \sum_{\lambda < \tau }
\sum_{i,j=1}^{4} \epsilon_{\mu \nu \lambda \tau } \, \mbox{Tr}
\left( p^{(i)}_{\mu \nu } \, p^{(j)}_{\lambda \tau } \right)
\label{topcharge}
\end{equation}
where $p^{(i)}_{\mu \nu } $, $i=1,\ldots ,4$ denotes the four
elementary lattice squares touching the lattice site $x$ and extending
into the $\mu $ and $\nu $ directions. The normalization of $q(x)$ can
be inferred by noting that vortex world-surface intersection points
generate contributions of magnitude $1/3$ or $2/3$ to the topological
charge \cite{contvort} (depending on the relative color orientation of the
surfaces). Note that, as written, each pair of elementary squares
is counted twice as $\mu , \nu , \lambda , \tau $ are summed over; this
is also properly taken into account by the normalization prefactor.

However, before a measurement of the topological charge according to
(\ref{topcharge}) can be implemented, certain ambiguities in the surface
configurations, engendered by the hypercubic construction, must first be
resolved.

\subsection{Ambiguities in measuring topological charge}
\label{ambisec}
\begin{figure}
\vspace{-0.3cm}
\centerline{
\epsfig{file=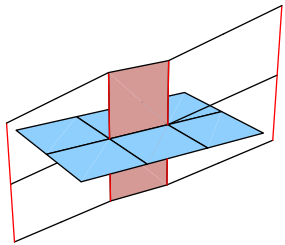,width=4cm}
}
\vspace{-0.8cm}
\caption{Vortex world-surfaces composed of elementary squares on a
hypercubic lattice exhibit ambiguities, such as two or more surface
segments meeting along a whole line, which do not arise in continuum
random surface ensembles.}
\label{intline}
\end{figure}

There are two types of ambiguities which arise in defining the topological
charge of hypercubic model surfaces. First, intersections of such surfaces
do not necessarily occur only at space-time points, but they can be spread
out into lines, as exemplified in Fig.~\ref{intline}. Contrast this with
the generic intersections found for arbitrary continuous two-dimensional
surfaces in four-dimensional space-time. In a generic ensemble of random
continuous surfaces, situations such as depicted in Fig.~\ref{intline},
where two or more surfaces meet along a whole line (or, in the lattice
language, where four or more elementary squares meet along a link)
represent a set of measure zero. Instead, intersections occur only at
points. In that case, one can unambiguously identify the two distinct
participating surfaces, and, moreover, they will each have a well-defined
orientation at the intersection, since monopole world-lines generally will
not run exactly through the intersection point. This leads to a well-defined
contribution to the topological charge.

By contrast, if an intersection point is spread out into a line, it can
happen that a monopole line crosses the intersection line, implying that
the surfaces intersecting do not possess unique orientations throughout the
intersection region. One might contemplate deforming the monopole line
around the intersection region, but it is unclear how to do so, because,
in general, it is not even possible to unambiguously distinguish between
the two surfaces which are intersecting in the first place. Given a
vortex configuration in which four or more elementary squares meet along
a link, there may be more than one way of assigning the squares in question
to two distinct surfaces; different assignments may even lead to different
conclusions as to whether an actual intersection point is observed or two
surfaces are merely touching.

This ambiguity in identifying the two distinct surfaces participating in
a situation where four or more vortex elementary squares meet along a
lattice link must be resolved before a topological charge can be assigned.
This is achieved by locally deforming the vortex world-surfaces until at
most three vortex elementary squares meet at each lattice link (the case
of three squares meeting, which does not occur in the $SU(2)$ model
\cite{m2}, constitutes a bona fide vortex branching allowed in the
$SU(3)$ case studied here). In practice, the given world-surface
configuration is placed on a finer lattice with $1/3$ of the original
lattice spacing, and one sweep is performed through the lattice, carrying
out elementary cube transformations, cf.~section~\ref{lattsurf}, such
as to eliminate lattice links with more than three vortex elementary
squares attached. This is quite efficient in practice; almost all
such ambiguities disappear on the first iteration of this algorithm,
and only one further iteration is necessary to completely eliminate
residual cases and arrive at an unambiguous surface configuration.
Thus, in practice, one ends up with a configuration on a lattice
with $1/9$ the original lattice spacing. Note that the algorithm only
performs local deformations in the sense that the original surfaces
are displaced by less than half of the original lattice spacing.
The values of Wilson loops defined in the original configuration
are thus manifestly unchanged. From the point of view of an infrared
effective model, the local deformations performed are smaller than
the uncertainty in defining vortex location implied by the ultraviolet
cutoff. However, while the confinement properties of the ensemble are
thus preserved, there is unfortunately the possibility of spurious
additional contributions to the topological charge being introduced on
the finer lattice scales (similar to the way instantons can ``fall
through the lattice'' in standard lattice Yang-Mills theory). The
corresponding systematic downward uncertainty in the measured topological
susceptibility will be estimated, cf.~section~\ref{ressec}, via the
simultaneous increase in vortex density caused by the algorithm, which
turns out to be appreciable. This will in fact represent the chief
uncertainty of the measurement.

The second ambiguity in hypercubic world-surface configurations which needs
to be removed is associated with the structure of vortices in color space.
In the hypercubic description, topological charge density is concentrated
at lattice sites; on the other hand, also magnetic monopole world-lines
are forced to run along lattice links, and, thus, through lattice sites.
The coincidence of a singular concentration of topological charge with
a magnetic monopole singularity is ambiguous. Contrast this again with
the generic structures found for arbitrary continuous two-dimensional
surfaces in four-dimensional space-time. In the case of an intersection
point, random monopole world-lines on the vortex surface will generally
pass by that exact point instead of going through it. In the case of
vortex writhe, which is continuously distributed along continuous
two-dimensional surfaces, a monopole world-line may indeed run through
the region of writhe; however, this has a negligible effect on the
topological charge density, for the following reason: Changing the
color orientation $T$ of a vortex surface segment leaves the topological
charge density generated by writhe within that segment invariant. Thus,
the only way in which a monopole world-line can influence topological charge
density is through interference of field strengths located on different
sides of the monopole. For thin surfaces, such interference is negligible;
for thickened surfaces, the situation is not quite as clear-cut,
cf.~further comments below. Disregarding for the moment the complications
implied by vortex thickness, the situation for continuous two-dimensional
surfaces is therefore this: Topological charge density generated by vortex
writhe is distributed continuously on the surface and insensitive to the
presence of monopoles, except exactly at the location of the monopole
world-line; however, when integrating the topological charge density,
the monopole world-line region, being of lower dimensionality, has
measure zero. This is different from the hypercubic case, where the
lattice description forces the entire topological charge density to be
concentrated at a lattice site, implying spurious finite interference
terms between field strengths located on either side of a monopole
world-line. To faithfully model the behavior of the topological charge
density of continuous vortex world-surfaces within the hypercubic
construction employed here, one should therefore deform all magnetic
monopole world-lines around lattice sites such that they never intersect
points of nonvanishing topological charge density. Note that this can be,
and is, effected locally and independently for each lattice site, once
one adopts the choice (\ref{nonmin}) for the description of the color
orientations possible for vortex world-surfaces. This is the point where
the adoption of the choice (\ref{nonmin}) is crucial; it permits the
resolution of the ambiguities associated with vortex color structure
within the hypercubic construction of the vortex world-surfaces employed
here. Further discussion of the choice of vortex color structure in the
dynamical configurations contained in the random vortex world-surface
ensemble follows below in section~\ref{colchoice}.

Returning to the case of thickened vortex world-surfaces deferred in the
discussion above, in general, one can indeed construct configurations in
which field strengths on different sides of a magnetic monopole may
interfere and thereby generate a contribution to the topological charge
density associated specifically with the monopole. An example of such
a configuration is given in \cite{cw3}. However, presumably, such
contributions are merely taken into account at a different level once
one adopts a construction in terms of thin world-surfaces. The connection
between thick vortices and their idealized thin representatives is
presumably topologically trivial, i.e., one can envisage continuously
deforming thick chromodynamic fluxes into thin constrictions thereof;
then, the topological charge density originally present in the thick flux
would remanifest itself in additional writhe and self-intersection of the
constricted flux. The further developments in the present work will base
on this presumption, whether it represents an auxiliary model assumption
or whether it indeed implies no loss of generality. Monopole lines will,
as described further above, always be deformed such that space-time
points with nonvanishing topological charge density are avoided; no
separate topological charge density will be associated with magnetic
monopole world-lines. In particular in the hypercubic description with
the choice of allowed color orientations (\ref{nonmin}), monopoles can
always be routed on the vortex world-surfaces such that all chromodynamic
flux in their immediate surroundings is confined to three dimensions,
i.e., cannot contribute to the topological charge density. It should be
emphasized that, within the present vortex model, magnetic monopoles are
not treated as separate physical degrees of freedom; rather, they are
merely manifestations of the non-orientedness of the vortex world-surfaces
arising in Abelian gauges such as implied by the choice (\ref{nonmin}).
Accordingly, their exact space-time location has no physical significance
to the extent that it can be varied by a change of gauge such that
vortex world-surface color orientation is rotated within the set
(\ref{nonmin}). While there are global constraints to such a change,
implied, e.g., by nonorientability, cf.~also a further discussion of
gauge invariance in section~\ref{colchoice}, the local deformations
necessary to remove all interactions between monopole world-lines and
space-time points carrying topological charge density are always
possible.

\subsection{Remarks on topological charge quantization}
\label{quantsec}
The global topological charge $Q$ of hypercubic model vortex world-surfaces
with $SU(3)$ color, evaluated as described above, is quantized in
half-integer units. The same property is exhibited in the $SU(2)$ case
\cite{m2}, cf.~also recent related work reporting evidence for half-integer
topological charge in a sample vortex configuration \cite{hellfab2}, as
well as the example given in \cite{m3}. To understand this behavior, it
should first be noted that vortex world-surfaces carrying global topological
charge are not defined on smooth, simple manifolds. In the Abelian gauge
language, a vortex configuration exhibiting nonvanishing global topological
charge must be non-oriented, i.e., carry Abelian magnetic monopoles
\cite{contvort}. The magnetic monopoles imply the presence of Dirac string
singularities in the vortex gauge field which must be excised from
space-time. As a consequence, the manifold supporting the vortex gauge field
acquires a complicated topology with internal boundaries, and topological
charge is not necessarily quantized in the manner which is found on
simple manifolds such as spheres or tori\footnote{Note that this is
not an artefact of the Abelian description; whereas one can indeed
construct (singular) non-Abelian gauge transformations which eliminate
Abelian magnetic monopoles and the associated Dirac strings \cite{contvort},
these transformations will not obey smooth boundary conditions at
the external boundaries of the manifold. The singular behavior is merely
shifted from internal boundaries in the region of the vortex carrying
topological charge to the external boundaries, thus precluding
compactification of the manifold at the latter to, say, a sphere or
a torus.}. Indeed, the topology of the space-time manifold is dynamic,
in close correspondence to the dynamic nature of the vortex topological
charge.

\begin{figure}
\vspace{-0.1cm}
\centerline{
\epsfig{file=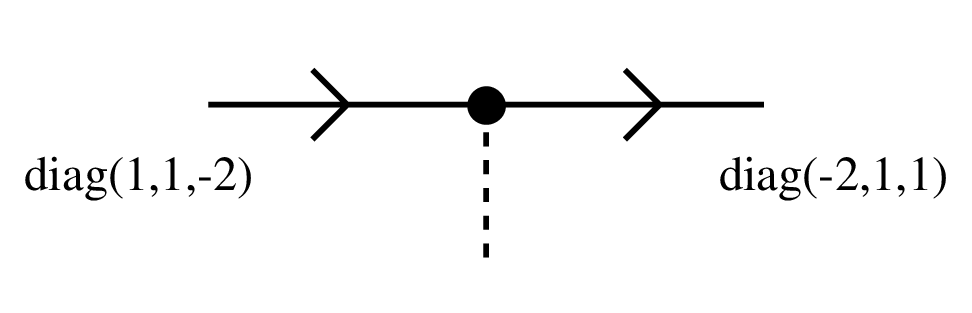,width=7cm}
\hspace{1.5cm}
\epsfig{file=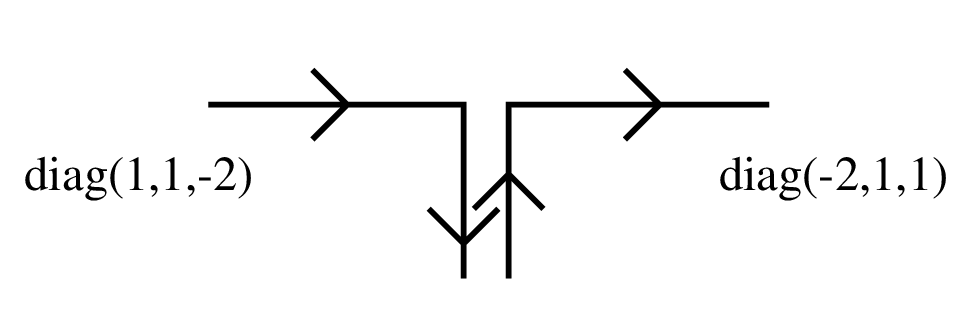,width=7cm}
}
\caption{Magnetic monopoles present in a generic vortex configuration
$A$ (left, where the broken line indicates the Dirac string) are replaced
by additional physical auxiliary fluxes in the corresponding configuration
$A^{\prime } $ (right). The auxiliary flux can furthermore be decomposed
into a superposition of coinciding vortex fluxes (depicted slightly
displaced from one another in the right-hand panel for better
visibility).}
\label{auxfig}
\end{figure}

To understand this correspondence in further detail, it is useful to
associate with any given Abelian vortex gauge field configuration $A$ a
corresponding configuration $A^{\prime } $ defined as follows. Let $A$ be
identical to $A^{\prime } $ everywhere except at Dirac strings. Instead of
excising Dirac string world-surfaces from space-time, let $A^{\prime } $
contain physical thin magnetic fluxes where $A$ exhibits Dirac strings,
cf.~Fig.~\ref{auxfig}. These magnetic fluxes will be referred to as
``auxiliary fluxes'' in the following. The magnitudes of these auxiliary
fluxes shall be multiples of $3$ (in each diagonal color component, in
the same convention as used for the color orientation matrices $T$
introduced in section~\ref{colstruc}), such as to supply precisely the
magnetic flux emanating from the magnetic monopoles in $A$.

The configuration $A^{\prime } $ thus has no sources or sinks of magnetic
flux, i.e., magnetic monopoles, and no Dirac strings. Instead, at any point
where vortex magnetic flux is discontinuous in $A$, in $A^{\prime } $, this
flux is continuously carried away by the newly introduced auxiliary fluxes.
Indeed, one can view all auxiliary fluxes as superpositions of additional
vortex fluxes with coinciding\footnote{Note that for the present argument,
auxiliary fluxes are not intended to be separated into non-coinciding
vortex world-surfaces. While this would indeed constitute yet another
legitimate point of view, it is not helpful in the present context,
since it in general introduces additional topologically nontrivial
features. For instance, in the slightly displaced depiction of
Fig.~\ref{auxfig} (right), consider connecting the two corners of the shown
fluxes by an imaginary line segment; adding another dimension, this
line segment becomes a band. Globally, this band may have the topology
of a M\"obius strip, which would imply additional writhe and
self-intersection in the separated vortex fluxes. Such complications
do not have to be taken into account if one foregoes separating the
auxiliary fluxes into disjoint center vortex units.} world-surfaces,
cf.~Fig.~\ref{auxfig}. Magnetic flux in $A^{\prime } $ is completely
continuous, and $A^{\prime } $ is defined on a manifold with no
internal boundaries. Consequently, its global topological charge
vanishes \cite{contvort}.

This opens the possibility of quantifying the topological charge of $A$
in an alternative manner, namely, via the properties of its Dirac strings,
or, equivalently, the properties of the corresponding auxiliary fluxes
in $A^{\prime } $. Consider the ways in which the topological charge
contributions found in $A^{\prime } $ differ from the ones found in $A$.
Besides the vortex topological charge proper, as measured in $A$,
$A^{\prime } $ contains the following additional contributions:
\begin{itemize}
\item Contributions from intersections between vortices and auxiliary
fluxes, $Q[A \cap Aux]$. These generate integer topological charge
contributions.
\item Contributions from intersections of auxiliary fluxes with auxiliary
fluxes, $Q_{int} [Aux]$. Also these generate integer topological charge
contributions.
\item Contributions from auxiliary flux writhe, $Q_{writhe} [Aux]$.
\end{itemize}
Note that, in accordance with the comments at the end of the previous
section, additional contributions from flux writhe at the edges of the
auxiliary fluxes, where the the original configuration $A$ displays
monopoles, are not contemplated. Monopoles are always routed such that
vortex flux in their immediate surroundings extends purely in three
dimensions, and one can convince oneself that attaching auxiliary
fluxes to lines routed in this fashion can also always be achieved
such that no writhe contributing to the topological charge density
results.

As a result of this construction, one can thus equate
\begin{equation}
0 = Q[A^{\prime } ] = Q[A] + Q[A \cap Aux]
+ Q_{int} [Aux] + Q_{writhe} [Aux] \ ,
\end{equation}
which implies that the quantization properties of $Q[A]$ are determined
by the quantization properties of $Q_{writhe} [Aux]$ (given that 
$Q[A \cap Aux]$ and $Q_{int} [Aux]$ are integers). In this sense,
there is a correspondence between the topology of the space-time manifold
and the quantization of the topological charge of the vortex world-surfaces.
Moreover, it thus becomes plausible that this quantization is independent
of the gauge group employed, as observed in practice and mentioned at the
beginning of this section\footnote{Note that the mechanism generating
fractional topological charge in the present context thus differs, e.g.,
from the topological charge fractionalization found using twisted boundary
conditions on a torus, which does depend on the gauge group
\cite{thootwist,baaltwist,df3}.}. After all, the magnitudes of the
auxiliary fluxes determining $Q_{writhe} [Aux]$ are independent of
the gauge group\footnote{Of course, for more than two colors, $N>2$,
more than one $SU(2)$ subgroup exists in which the gauge field can exhibit
an Abelian magnetic monopole and the associated Dirac string; however,
any given Dirac string in any specific $SU(2)$ subgroup of an $SU(N)$
group carries the same quantum of magnetic flux, independent of $N$.},
and consequently, it is plausible that also the quantization properties
of $Q_{writhe} [Aux]$ (and, therefore, $Q[A]$) would be independent of
the gauge group.

A caveat to this argument is the following: Whereas locally, the
contributions to $Q_{writhe} [Aux]$ stemming from writhe and
self-intersection of the auxiliary fluxes are indeed independent
of the gauge group, globally, the topologies of the auxiliary
flux world-surfaces differ. E.g., for $SU(3)$, auxiliary flux may
branch, whereas it cannot for $SU(2)$. Now, in complete analogy to
the argument used further above, excluding auxiliary flux writhe at the
edges where auxiliary flux is attached to the physical vortices, one can
also always configure $SU(3)$ auxiliary flux branching lines such that
they do not carry writhe. Thus, both for $SU(2)$ and for $SU(3)$,
$Q_{writhe} [Aux]$ is determined by writhe and self-intersections in the
interior of open auxiliary flux world-surfaces, with the only difference
that, in the $SU(3)$ case, the edges of the open world-surfaces include
not only the lines where the auxiliary flux is attached to the physical
vortices, but may also include auxiliary flux branching lines. It thus
remains plausible that $Q_{writhe} [Aux]$ is quantized in the same way
for both gauge groups. Nevertheless, at this point there is no stringent
argument excluding the possibility that global space-time constraints may
induce differences in the available sets of open auxiliary flux
world-surfaces for the two gauge groups. As a result, the above
observations, motivating the coinciding topological charge quantization
properties of $SU(2)$ and $SU(3)$ model center vortices, should be taken
as no more than an a posteriori plausibility argument.

\section{Vortex ensemble}
\subsection{Dynamics}
\label{mcdyn}
The dynamics of the $SU(3)$ random vortex world-surface model were
constructed and discussed in detail in \cite{su3conf}. Vortex world-surfaces
are composed of elementary squares on a hypercubic lattice, as described
in section~\ref{lattsurf}. An ensemble of random vortex world-surfaces
is generated via Monte Carlo methods, where preservation of the Bianchi
constraint (continuity of flux modulo Abelian magnetic monopoles) is
guaranteed by using the elementary cube transformations of
section~\ref{lattsurf} as the elementary updates. The ensemble is weighted
by an action penalizing curvature of vortex surfaces. In terms of the
reduced quantities\footnote{Note that previous discussions of the
$SU(3)$ random vortex world-surface model \cite{su3conf,su3bary,su3freee}
were formulated directly in terms of the reduced variables
$q_{\mu \nu } (x)$, since these are sufficient for encoding the values
taken by Wilson loops in vortex configurations, and thus sufficient
for the discussion of confinement properties, on which
\cite{su3conf,su3bary,su3freee} focus. By contrast, for the discussion
of topological properties, it is useful to introduce the more detailed
specification of color orientation provided by the variables
$p_{\mu \nu } (x)$ used in the present work. The space-time dynamics
of the ensemble, depending only on the absolute values $|q_{\mu \nu } (x)|$,
are unchanged.}
\begin{equation}
q_{\mu \nu } (x) = \mbox{sgn} \, \, \mbox{Im} \, \, \mbox{Tr} 
\exp (2\pi i\, p_{\mu \nu } (x) /3) \ ,
\label{qred}
\end{equation}
i.e., $q_{\mu \nu } (x) \in \{-1,0,1\} $, the action is
\begin{eqnarray}
S &=&
c \sum_x\sum_\mu \left[ \sum_{\nu < \lambda \atop \nu \neq \mu,
\lambda\neq \mu} \left( | q_{\mu\nu}(x) \, q_{\mu\lambda}(x) |
 + | q_{\mu\nu}(x) \, q_{\mu\lambda}(x-e_\lambda) |
\right. \right. \label{curvature} \\
& & \ \ \ \ \ \ \ \ \ \ \ \ \ \ \ \ \ \
+ \left. | q_{\mu\nu}(x-e_\nu) \, q_{\mu\lambda}(x) |
 + | q_{\mu\nu}(x-e_\nu) \, q_{\mu\lambda}(x-e_\lambda) |
\right)\Bigg] \ .
\nonumber
\end{eqnarray}
Thus, for every link in the lattice, the attached elementary squares
carrying vortex flux are considered, and every instance of a pair of these
squares not lying in the same plane costs an action increment $c$. The
value of $c$,
\begin{equation}
c=0.21
\end{equation}
is fixed \cite{su3conf} by demanding that the ratio of the deconfinement
temperature to the square root of the zero temperature string tension
reproduce the value obtained in $SU(3)$ Yang-Mills theory \cite{boyd},
$T_c /\sqrt{\sigma } =0.63$.

Note, finally, that the lattice spacing is a fixed physical quantity,
encoding the ultraviolet cutoff of this infrared effective model;
physically, it implements the notion that vortices possess a finite
transverse thickness. E.g., a pair of parallel thick vortices can only
approach one another up to a minimal distance, below which their fluxes
cease to be distinguishable from one another and should instead be
represented as one combined flux with an appropriate new color
orientation. It is therefore not meaningful to consider configurations
in which two vortices run in parallel at a distance smaller than the
aforementioned minimal one. This is reflected in the fixed lattice
spacing used in conjunction with the hypercubic construction of the
vortex world-surfaces. Fixing the scale by setting the zero-temperature
string tension to $\sigma =(440\, \mbox{MeV})^{2} $, the lattice spacing
takes the value $a=0.39\, \mbox{fm} $, cf.~\cite{su3conf}.

\subsection{Color structure}
\label{colchoice}
The action (\ref{curvature}) contains no bias with respect to
the color orientation of vortex flux. Also, the Bianchi identity only
constrains the reduced quantities $q_{\mu \nu } (x)$, but does not
distinguish between different color orientations of $p_{\mu \nu } (x)$
corresponding to the same $q_{\mu \nu } (x)$. This is in accordance with
the fact that the color orientation of the (vortex) field strength
can be locally rotated by gauge transformations in the underlying
full Yang-Mills theory\footnote{Note that, in general, not all
$p_{\mu \nu } (x)$ corresponding to the same $q_{\mu \nu } (x)$
are related by gauge transformations; this point will be revisited in more
detail presently. Thus, the dynamics embodied in (\ref{curvature}) are
invariant not only under bona fide gauge transformations, but under a
larger class of transformations connecting all possible color orientations
of the vortex field strength.}.

Nevertheless, it is necessary to make a specific choice of color orientation,
i.e., of the full quantities $p_{\mu \nu } (x)$ for the purpose of evaluating
the topological charge according to (\ref{topcharge}). In practice, vortex
configurations are generated as in previous investigations of the $SU(3)$
random vortex world-surface model \cite{su3conf,su3bary,su3freee}, i.e.,
in terms of $q_{\mu \nu } (x)$. Then, all vortex elementary squares are
assigned random color orientations from the allowed set of $T$,
cf.~(\ref{nonmin}), consistent with the given $q_{\mu \nu } (x)$, thus
arriving at an initial model description in terms of the full quantities
$p_{\mu \nu } (x)$.

Before continuing, it should be emphasized that this assignment is not
necessarily just a particular choice of gauge. Certainly, smooth
local color rotations of the (vortex) field strength amount to gauge
transformations, and one might therefore be tempted to regard a choice
of $p_{\mu \nu } (x)$ for a given $q_{\mu \nu } (x)$ purely as a choice
of gauge. However, while different choices of color orientation indeed
fall into gauge equivalence classes, there is, in general, more than
one class. Gauge-inequivalent choices are possible, in particular, at
vortex intersection points: If one assigns color orientations to vortex
world-surface elements completely independently, this also allows one to
change the color orientation of one, but not the other, vortex surface
meeting at the intersection point in question. This goes beyond what is
possible using gauge transformations, which only allow one to rotate
the entire gauge field strength present at a given point coherently.
Thus, the assignment of color orientation to the vortex world-surfaces is
related to, but not synonymous with a choice of gauge. Gauge-inequivalent
$p_{\mu \nu } (x)$ for a given $q_{\mu \nu } (x)$ are possible, and these,
in general, also differ in their topological charge. The topological charge
can vary within certain bounds depending on the color orientation
chosen for the configuration; it is not determined exclusively by the
space-time location of the vortex world-surfaces.

It is thus certainly a relevant question to what extent the measurement
of the topological susceptibility is biased by the way vortex color
orientation is modeled. In order to glean some information regarding this
issue, in practice, two alternative models are considered: Starting with
the initial random assignment of color orientation in $p_{\mu \nu } (x)$
described above, sweeps through the lattice are performed in which the
color orientations of individual elementary squares are changed with a
bias towards either aligning or not aligning the orientations of
adjacent squares; this corresponds to minimizing or maximizing the
Abelian magnetic monopole density, respectively\footnote{The, in general,
nonorientable nature of the vortex world-surfaces implies a lower bound
on the monopole density achievable in this way, while the lattice spacing
provides an upper cutoff.}. In this way, one
arrives at alternative ensembles comprised of configurations
$p_{\mu \nu }^{min} (x)$ and $p_{\mu \nu }^{max} (x)$, on both of
which topological charge is measured (after removing the lattice
ambiguities discussed in section~\ref{ambisec}). While the two ensembles
will be seen to differ widely in monopole density, the results obtained in
either case for the topological susceptibility exhibit only minor
deviations from each other. To this extent, thus, an unambiguous
prediction of the topological susceptibility emerges despite the
freedom one has in modeling vortex color orientation. The reason for
this lies in the fact that the topological charge of generic vortex
world-surfaces is dominated by vortex writhe, as will be seen in more detail
in section~\ref{ressec}. Vortex intersections, at which gauge-inequivalent
color orientation choices leading to differing topological charge are
possible, cf.~the discussion further above, by contrast only generate a
minor contribution to the overall vortex topological charge. Note that
entirely analogous observations were already made in the $SU(2)$ random
vortex world-surface model, cf.~\cite{m2}.

\section{Numerical results and discussion}
\subsection{Measurement parameters}
Measurements were carried out at the physical value of the curvature
coefficient, $c=0.21$, on $12^3 \times N_t $ lattices, variation of $N_t $
permitting the study of a collection of temperatures including both the
confined as well as the deconfined phases. Furthermore, in order to
obtain more closely spaced data as a function of temperature than
provided by these direct measurements at $c=0.21$, the following
interpolation procedure was used in addition: A determination
\cite{su3freee} of the critical values of $c$ at which the deconfinement
transition occurs for $N_t =1,2,3$, cf.~Table~\ref{critc}, yields
\begin{table}[h]
\[
\begin{tabular}{|c||c|c|c|}
\hline
$N_t $  & 1 & 2 & 3 \\\hline\hline
$c$ & 0.0872 & 0.2359  & 0.335  \\\hline
\end{tabular}
\]
\caption{Critical values of the curvature coefficient $c$ at which
the deconfining phase transition occurs.}
\label{critc}
\end{table}
$aT_c = 1/N_t $ for those values of $c$. An interpolating parabola in $c$
then defines $aT_c $ for a continuous range of $c$ (in particular,
$aT_c = 0.5655$ for $c=0.21$, i.e., a $N_t =1$ lattice corresponds to
$T/T_c = 1.77$ at the physical value of $c$). On this basis, then,
$T/T_c = 1/(N_t \cdot aT_c (c))$ is defined for any combination of
$c, N_t $. Now, one can perform measurements at a given fixed value
of $T/T_c $ for different $N_t $ and the corresponding $c$. Interpolating
the results as a function of $c$ to the physical point $c=0.21$ finally
yields the desired supplementary data at the chosen $T/T_c $.
Table~\ref{cnt} lists the additional combinations of $N_t $ and $c$
at which measurements were
\begin{table}[h]
\[
\begin{tabular}{|c||c|c|c|}
\hline
$T/T_c $  & $N_t =1$ & $N_t =2$ & $N_t =3$ \\\hline\hline
0.98 & 0.082565 & 0.23162  & 0.32854  \\\hline
1.02 & 0.09172 & 0.24012  & 0.3418  \\\hline
1.1 & 0.10872 & 0.25642  & -- \\\hline
1.4 & 0.16102 & 0.3143  & -- \\\hline
\end{tabular}
\]
\caption{Temperatures and corresponding curvature coefficients $c$ at which
measurements were performed for $N_t =1,2,3$, to supplement the direct
measurements at $c=0.21$.}
\label{cnt}
\end{table}
performed for use in the above interpolation procedure; note that only
values of $c$ inside, or very close to the range covered by the
critical values listed in Table~\ref{critc} were used, in order to
preclude extrapolation instabilities. As a consequence, the final
interpolation to $c=0.21$ relies on three data points in the cases
of $T/T_c =0.98$ and $T/T_c =1.02$, whereas it relies on two data points
in the other two cases.

\subsection{Numerical results}
\label{ressec}
\begin{table}[h]
\[
\begin{tabular}{|c||c|c|c|c|}
\hline
$T/T_c $  & $\chi^{1/4} /\mbox{MeV} $ & $\chi^{1/4}_{int} /\mbox{MeV} $ &
$\rho_{monop} \cdot \mbox{fm}^{3} $ &
$\rho_{def} /\rho_{orig} $ \\\hline\hline
0.15 & 222 & 98 & 12.6 & 1.39 \\\hline
0.29 & 222 & 98 & 12.6 & 1.39 \\\hline
0.44 & 222 & 98 & 12.6 & 1.39 \\\hline
0.59 & 222 & 98 & 12.5 & 1.39 \\\hline
0.88 & 221 & 100 & 11.9 & 1.39 \\\hline
0.98 & 222 & 105 & 11.0 & 1.40 \\\hline
1.02 & 209 & 102 & 9.41 & 1.38 \\\hline
1.1 & 189 & 99 & 6.42 & 1.39 \\\hline
1.4 & 156 & 92 & 3.25 & 1.43 \\\hline
1.77 & 150 & 93 & 1.96 & 1.47 \\\hline
\end{tabular}
\]
\caption{Numerical results for ensemble with minimized magnetic monopole
density. Statistical uncertainties are smaller than the accuracy to which
quantities are quoted. The various columns are explained in detail in the
main text.}
\label{resmin}
\end{table}
\begin{table}[h]
\[
\begin{tabular}{|c||c|c|c|c|}
\hline
$T/T_c $  & $\chi^{1/4} /\mbox{MeV} $ & $\chi^{1/4}_{int} /\mbox{MeV} $ &
$\rho_{monop} \cdot \mbox{fm}^{3} $ &
$\rho_{def} /\rho_{orig} $ \\\hline\hline
0.15 & 224 & 98 & 63.2 & 1.39 \\\hline
0.29 & 223 & 98 & 63.2 & 1.39 \\\hline
0.44 & 223 & 98 & 63.2 & 1.39 \\\hline
0.59 & 223 & 99 & 63.2 & 1.39 \\\hline
0.88 & 223 & 101 & 63.4 & 1.39 \\\hline
0.98 & 223 & 107 & 61.7 & 1.40 \\\hline
1.02 & 210 & 105 & 58.0 & 1.38 \\\hline
1.1 & 191 & 104 & 47.1 & 1.39 \\\hline
1.4 & 161 & 98 & 30.9 & 1.43 \\\hline
1.77 & 156 & 98 & 23.5 & 1.47 \\\hline
\end{tabular}
\]
\caption{Numerical results for ensemble with maximized magnetic monopole
density. Statistical uncertainties are smaller than the accuracy to which
quantities are quoted. The various columns are explained in detail in the
main text.}
\label{resmax}
\end{table}
The results for the topological susceptibility
$\chi = \langle Q^2 \rangle / V$ (where $V$ denotes the space-time
four-volume) are given in Tables \ref{resmin} and \ref{resmax}, the
former referring to the ensemble with minimized Abelian monopole density,
cf.~the discussion in section~\ref{colchoice}, the latter referring to the
ensemble with maximized Abelian monopole density. For both cases,
besides the total topological susceptibility $\chi $, also the susceptibility
$\chi_{int} $ resulting from counting only the contributions from vortex
world-surface intersection points\footnote{Note that counting only
contributions from intersection points leads to a topological charge
$Q_{int} $ quantized in units of $1/3$.} is quoted. Furthermore, the
Abelian magnetic monopole density $\rho_{monop} = L_{monop} /V$
(evaluated before applying the algorithms removing the ambiguities
discussed in section~\ref{ambisec}) is given, where $L_{monop} $ is the
monopole world-line length present in the configuration (normalized such
that a single monopole world-line occupying a lattice link contributes a
length $a$). Finally, the aforementioned algorithms, removing hypercubic
lattice surface ambiguities in order to make an unambiguous evaluation of
the topological charge possible, cf.~section~\ref{ambisec}, involve
local deformations of the vortex world-surfaces; this, however, leads
to an appreciable increase in the vortex world-surface density.
The ratio of the deformed world-surface density $\rho_{def} $ to
the original density $\rho_{orig} $ is reported in the final column in
Tables~\ref{resmin} and \ref{resmax}; these data are the same in the two
tables, since the ensembles in question differ exclusively in the choice
of color structure, whereas the space-time locations of the vortices are
identical.

Evidently, the ensembles with minimized and maximized monopole
densities, respectively, differ considerably in color structure, as
quantified by those densities. Yet, the differences in the topological
susceptibilities measured in the two cases are minor, increasing
slightly at high temperatures. To this extent, the prediction for
the topological susceptibility obtained in the $SU(3)$ random vortex
world-surface model is unambiguous as far as the modeling of vortex
color structure is concerned. The reason for this behavior lies
in the space-time structure of generic world-surface configurations.
As seen in Tables \ref{resmin} and \ref{resmax}, the topological
susceptibility induced by world-surface intersection points alone is much
smaller than the contribution from vortex writhe; even when considering the
fourth root, $\chi^{1/4} $ and $\chi_{int}^{1/4} $ still differ roughly by
a factor of two. Vortex writhe is the dominant mechanism by which center
vortices generate topological charge. However, the contribution from
vortex writhe is explicitly invariant under color rotations of the
vortex field strength, since only one world-surface is involved in
generating such contributions. Only vortex intersection points involve
two distinct surfaces, independent color rotations of which can lead
to gauge-inequivalent color configurations, and thus change topological
charge. The relative paucity of such intersection points in generic
world-surface configurations, as evidenced by the magnitude of
$\chi_{int} $, explains the very similar results for the topological
susceptibility obtained in the two ensembles investigated, despite
their considerably differing color structure.

\begin{figure}
\centerline{
\epsfig{file=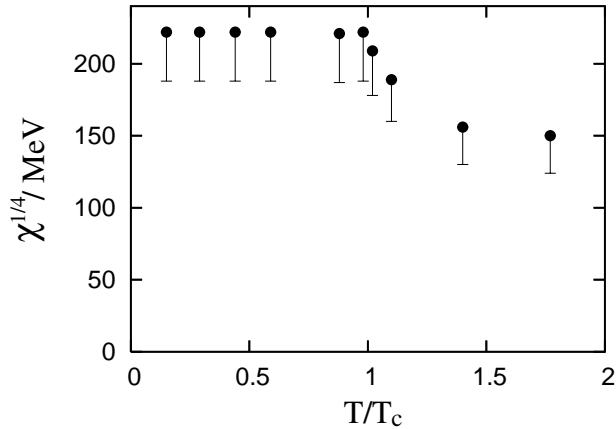,width=6cm,angle=-90}
}
\caption{Fourth root of the topological susceptibility measured in the
$SU(3)$ random vortex world-surface model, as a function of temperature.
The results from the ensemble with minimized Abelian magnetic monopole
density are depicted. Statistical uncertainties of the data are smaller
than the filled circle symbols displaying the measured values;
the downward uncertainty shown is a systematic one, discussed in detail
in the main text.}
\label{chifig}
\end{figure}

The results for the topological susceptibility summarized in Tables
\ref{resmin} and \ref{resmax} suffer from one significant systematic
uncertainty. Namely, as discussed in section~\ref{ambisec}, an unambiguous
evaluation of the topological charge of a hypercubic vortex world-surface
configuration only becomes possible after an algorithm is applied during
which the configurations are placed on finer lattices and subjected to
suitable local deformations. These deformations appreciably increase the
vortex density, as evidenced by the ratio $\rho_{def} /\rho_{orig} $
reported in Tables~\ref{resmin} and \ref{resmax}. In general, in the
process, also spurious additional topological charge density will be
generated concomitantly on the finer lattices. To obtain a rough estimate
of this effect, by simply counting dimensions, the topological
susceptibility would be expected to scale with the square of the vortex
world-surface density (a critique of this estimate is given below). On this
basis, an estimate for the amount by which the raw data for the fourth root
of the topological susceptibility given in Tables~\ref{resmin} and
\ref{resmax} may need to be systematically revised downward, to offset
the effects of the deformation procedure, can be obtained by dividing those
data by the square root of the ratio $\rho_{def} /\rho_{orig} $. This
yields the lower ends of the error bars depicted in Fig.~\ref{chifig},
which, for definiteness, shows the results obtained from the ensemble
with minimized magnetic monopole density.

One may alternatively contemplate interpreting the lower ends of the error
bars in Fig.~\ref{chifig} not merely as estimates of a systematic uncertainty,
but as appropriately adjusted data in their own right, akin to renormalized
quantities. Of course, this is mainly a metaphorical interpretation, since
no systematic scheme has been developed within the present infrared effective
model which would allow one to quantify the dependence on the lattice
spacing. However, on the finer lattices on which the deformations of the
vortex world-surfaces and the ensuing evaluation of their topological charge
are performed, presumably a suitably renormalized effective action exists
which would directly generate the deformed world-surfaces if one worked from
the very beginning on those finer lattices. The measurement of the topological
susceptibility is carried out on the finer lattice, in the ensemble controlled
by the corresponding renormalized action. Thus, it seems plausible that
translating the measurement back to the original scale should be accompanied
by a suitable renormalization, and that it is in principle appropriate to
discuss the results in terms of corresponding renormalized quantities. Of
course, the rescaling by the density ratio $\rho_{def} /\rho_{orig} $
employed here is no more than a rough phenomenological estimate of this
renormalization; after all, the fluctuations of the vortex world-surfaces
engendered by their local deformation on the finer lattices, while to a
certain extent random, are not governed by a simple curvature action of
the form (\ref{curvature}), and therefore one must expect corrections to
the simple scaling with the vortex density. These caveats must be kept in
mind in the context of interpreting the rescaled topological susceptibility
as a physical quantity in its own right.

Comparing to the $SU(2)$ random vortex world-surface model studied in
\cite{m2}, the modifications of the world-surfaces in the course of the
deformation algorithm, as quantified by the density ratio
$\rho_{def} /\rho_{orig} $, become stronger as one progresses to $SU(3)$.
At low temperatures, $\rho_{def} /\rho_{orig} =1.39$ in the $SU(3)$ model,
whereas one has only $\rho_{def} /\rho_{orig} =1.19$ for $SU(2)$. Similarly,
in the deconfined phase, the ratio rises to $\rho_{def} /\rho_{orig} =1.47$
at $T=1.77\, T_c $ in the $SU(3)$ model, whereas, in the $SU(2)$ model
at $T=1.66\, T_c $, one finds $\rho_{def} /\rho_{orig} =1.20$.
This behavior is plausible in view of the fact that $SU(3)$ vortices
each carry less chromomagnetic flux than $SU(2)$ vortices. As a consequence,
a given amount of flux will be fragmented into more world-surface area
in the $SU(3)$ case. Indeed, this is borne out by the data: In the $SU(2)$
model, at low temperatures, $27\% $ of elementary squares in the lattice
carry vortex flux (on the original coarse lattice, before deformation);
by contrast, in the $SU(3)$ model, this rises to $36\% $ of elementary
squares. It seems plausible that, the higher the percentage of occupied
squares, the more elaborate the deformations necessary to eliminate all
world-surface ambiguities become. Hence, the enhanced
$\rho_{def} /\rho_{orig} $ ratio for $SU(3)$ as compared to $SU(2)$.

In terms of the topological susceptibility\footnote{For definiteness, the
discussion in the remainder of this section and also in the next section
will refer specifically to the results obtained in the ensemble with
minimized Abelian magnetic monopole density, cf.~Table~\ref{resmin}.},
the raw data for the $SU(3)$ vortex model are considerably higher than
for the $SU(2)$ model\footnote{The comparison between $SU(3)$ and
$SU(2)$ is performed on the basis of positing an identical zero-temperature
string tension, $\sigma (T=0) = (440\, \mbox{MeV} )^2 $, to set the scale
in both cases.} investigated in \cite{m2}; namely, at low temperatures,
$\chi_{raw}^{1/4} = 222\, \mbox{MeV} $ for $SU(3)$
vs.~$\chi_{raw}^{1/4} = 190\, \mbox{MeV} $ for $SU(2)$. Here and in the
following, $\chi_{raw}^{1/4} $ simply denotes the measured data labeled
as $\chi^{1/4} $ in Table~\ref{resmin} and displayed by the filled circles
in Fig.~\ref{chifig}; on the other hand, the data rescaled with
$\rho_{def} /\rho_{orig} $ as discussed above, corresponding to the lower
ends of the error bars displayed in Fig.~\ref{chifig}, will be denoted by
$\chi_{rescaled}^{1/4} $, i.e., $\chi_{rescaled}^{1/4} = \chi_{raw}^{1/4} /
\sqrt{\rho_{def} /\rho_{orig} } $. In terms of the rescaled quantities,
the comparison between the $SU(3)$ and $SU(2)$ models at low temperatures
is much closer, although the $SU(3)$ result still lies above the $SU(2)$ one,
namely, $\chi_{rescaled}^{1/4} = 188\, \mbox{MeV} $
vs.~$\chi_{rescaled}^{1/4} = 174\, \mbox{MeV} $. At high temperatures,
the contrast is stronger: In the $SU(3)$ model at $T=1.77\, T_c $, one
has $\chi_{raw}^{1/4} = 150\, \mbox{MeV} $, whereas the $SU(2)$ model
at $T=1.66\, T_c $ yields $\chi_{raw}^{1/4} = 109\, \mbox{MeV} $. On the
other hand, in terms of the rescaled quantities at the same temperatures,
$\chi_{rescaled}^{1/4} = 124\, \mbox{MeV} $ for $SU(3)$, while
$\chi_{rescaled}^{1/4} = 100\, \mbox{MeV} $ for $SU(2)$. These
comparisons will be revisited below in relation to corresponding
lattice Yang-Mills results.

\subsection{Comparison to lattice Yang-Mills theory}
Lattice Yang-Mills results for the $SU(3)$ topological susceptibility
have been reported in a number of works, cf., e.g.,
\cite{allessu3,gattr,ltw,edwards,lt01,cundy,forchet,allessu2,deldebb,hoelb},
and reviewed in \cite{panvic}; cf.~the latter also for a much more
extensive list of related studies. There is a considerable spread in
the reported results, obtained using various methods, with values as high as
$\chi^{1/4} = 213(7)\, \mbox{MeV} $ and as low as
$\chi^{1/4} = 168(11)\, \mbox{MeV} $ obtained at zero temperature in the
last decade, cf.~Table~1 in \cite{panvic}. The corresponding raw data at
low temperatures found in the $SU(3)$ random vortex world-surface model
lie somewhat above this range, at $\chi_{raw}^{1/4} = 222\, \mbox{MeV} $.
On the other hand, the rescaled result,
$\chi_{rescaled}^{1/4} = 188\, \mbox{MeV} $, lies near the center of the
range of lattice Yang-Mills data; the vortex model results thus appear
compatible with the lattice Yang-Mills results at low temperatures.

Continuing to finite temperature, above the deconfining phase transition,
the topological susceptibility is seen to fall off rapidly with temperature
in $SU(3)$ lattice Yang-Mills theory \cite{allessu3,gattr,ltw,edwards}.
Quantitatively, \cite{allessu3} reports a drop in $\chi^{1/4} $ by a
factor $2.3$ as the temperature is raised from $T=0.834\, T_c $ to
$T=1.402\, T_c $; \cite{gattr} reports a drop in $\chi^{1/4} $ by a
factor $1.9$ as the temperature is raised from $T=0.88\, T_c $ to
$T=1.31\, T_c $. An even stronger suppression is reported by \cite{ltw},
namely, by a factor $2.8$ as the temperature rises from just below the
deconfinement temperature, $T=0.99\, T_c $, up to $T=1.25\, T_c $. This seems
significantly different from the former two measurements, which appear
compatible with each other. Finally, \cite{edwards} gives only data in
the deconfined phase, reporting a drop in $\chi^{1/4} $ by a factor
$1.9$ as the temperature is raised from just above the transition,
$T=1.03\, T_c $, to $T=1.38\, T_c $ (using the data obtained on
$16^3 \times 4$ lattices quoted in \cite{edwards}). If one combines
this with data on the discontinuity across the deconfining phase
transition \cite{ltw}, indicating a drop in $\chi^{1/4} $ by an additional
factor $1.15$ across the transition, this cumulatively amounts to a drop
in $\chi^{1/4} $ by a factor $2.2$ as one increases the temperature from
the confined phase up to $T=1.38\, T_c $. This again seems compatible with
the results from \cite{allessu3,gattr}.

Also in the $SU(3)$ random vortex world-surface model, the topological
susceptibility quickly becomes suppressed in the deconfined phase as
temperature rises, cf.~Fig.~\ref{chifig}; however, quantitatively, the
suppression is not as strong as the one seen in $SU(3)$ lattice
Yang-Mills theory. The raw topological susceptibility data in
Table~\ref{resmin} show a drop in $\chi_{raw}^{1/4} $ from
$\chi_{raw}^{1/4} = 221\, \mbox{MeV} $ to
$\chi_{raw}^{1/4} = 156\, \mbox{MeV} $, i.e., by a factor $1.42$,
as the temperature is raised from $T=0.88\, T_c $ to $T=1.4\, T_c $.
This does not change substantially when the rescaled data are considered,
since the density ratio $\rho_{def} /\rho_{orig} $ does not vary strongly
with temperature; in terms of rescaled data, $\chi_{rescaled}^{1/4} $
drops by a factor $1.44$ in the same temperature range. Thus, the
topological susceptibility found in the $SU(3)$ random vortex world-surface
model in the deconfined phase does appear to remain appreciably above
the corresponding lattice Yang-Mills results. The comparison with
lattice Yang-Mills theory in the present $SU(3)$ case in the deconfined
phase therefore is less favorable than for the previously studied
$SU(2)$ model \cite{m2}, which is quantitatively compatible with
corresponding lattice Yang-Mills results even above the deconfining
transition. Possible causes of this will be discussed in the next
section.

Another way to cast the comparison between the random vortex world-surface
model and lattice Yang-Mills theory is in terms of the trend, already
alluded to at the end of the previous section, as one progresses from
the $SU(2)$ to the $SU(3)$ gauge group. This in fact yields the starkest
contrast. In lattice Yang-Mills theory, at low temperatures, the
topological susceptibility is expected to fall as the number of colors $N$
is increased \cite{panvic}. The preponderance of evidence points to this
already being the case as one goes from $SU(2)$ to $SU(3)$
\cite{lt01,ltw,cundy,forchet}, although it should be noted that
agreement on this only emerges when the results are extrapolated to
vanishing lattice spacing \cite{lt01,forchet}; at finite lattice
spacing, a slight increase of the zero-temperature topological
susceptibility going from $SU(2)$ to $SU(3)$ has been seen \cite{lt01,ltw}.
A detailed discussion of the continuum extrapolation can be found
in \cite{lt01}. Disregarding these details, in the confined phase,
the lattice Yang-Mills topological susceptibilities found in the
$SU(2)$ and $SU(3)$ cases only differ mildly. In this respect, the
random vortex world-surface model can still be viewed as compatible with
lattice Yang-Mills theory; as already observed at the end of the previous
section, while the raw topological susceptibility data obtained at
low temperatures in the $SU(3)$ and $SU(2)$ cases differ substantially,
$\chi_{raw}^{1/4} = 222\, \mbox{MeV} $ for $SU(3)$
vs.~$\chi_{raw}^{1/4} = 190\, \mbox{MeV} $ for $SU(2)$, in terms of
the rescaled quantities, the $SU(3)$ result only lies slightly above
the $SU(2)$ result, namely, $\chi_{rescaled}^{1/4} = 188\, \mbox{MeV} $
vs.~$\chi_{rescaled}^{1/4} = 174\, \mbox{MeV} $.

The picture changes qualitatively as one crosses into the deconfined phase.
While the vortex model does display a strong drop in the topological
susceptibility both for $SU(3)$ and $SU(2)$ as the temperature is
raised, as already discussed further above, the suppression seen is
substantially stronger for $SU(2)$ than for $SU(3)$, even in terms
of the rescaled quantities. In the $SU(3)$ model at $T=1.77\, T_c $,
one has $\chi_{rescaled}^{1/4} = 124\, \mbox{MeV} $, whereas already
at $T=1.66\, T_c $, the $SU(2)$ model yields
$\chi_{rescaled}^{1/4} = 100\, \mbox{MeV} $. This does
seem clearly at odds with the behavior seen in lattice Yang-Mills
theory. There, all evidence points towards the reverse
relation\footnote{Indeed, one expects the topological susceptibility
to vanish in the deconfined phase as $N\rightarrow \infty $ \cite{panvic}.}:
\cite{allessu2} shows (cf.~Fig.~4 therein), at $T=1.3\, T_c $, a $SU(3)$
topological susceptibility which is smaller by roughly a factor $10$
compared to the $SU(2)$ susceptibility; in terms of the fourth root,
$\chi^{1/4} $, this means a suppression of the $SU(3)$ result compared
to the $SU(2)$ result by a factor $1.8$ at $T=1.3\, T_c $. Also
\cite{edwards} reports data for both $SU(3)$ at $T=1.38\, T_c $, and
for $SU(2)$ at $T=1.4\, T_c $. The data are in lattice units; using
the results obtained in \cite{edwards} at the aforementioned temperatures
on $16^3 \times 4$ lattices, one has
$(\chi [SU2]/\chi [SU3]) \cdot (a[SU(2)]/a[SU(3)])^4 = 5.1$ at roughly
$T=1.4\, T_c $. One can convert to physical units, e.g., by observing
that $T_c a$ is roughly the same in both measurements; then, using
$T_c /\sqrt{\sigma} =0.69$ for $SU(2)$ and $T_c /\sqrt{\sigma} =0.63$
for $SU(3)$ ($\sigma $ denoting the zero-temperature string tension),
this implies that $a[SU(2)]/a[SU(3)] = 0.9$. Taken together, therefore,
at roughly $T=1.4\, T_c $, the $SU(3)$ result for $\chi^{1/4} $ is
suppressed compared to the $SU(2)$ result by a factor $1.67$, similar
to the behavior shown in \cite{allessu2}. Thus, in the deconfined phase,
substantial disagreement is seen between the random vortex world-surface
model and lattice Yang-Mills theory, as far the comparison between the
topological susceptibilities obtained in the $SU(3)$ and $SU(2)$ cases is
concerned.

\section{Outlook}
The results presented and discussed in the previous section indicate
that the topological properties of the $SU(3)$ random vortex world-surface
model investigated in this work are consistent with $SU(3)$ Yang-Mills
theory in the confined phase. However, in the deconfined phase, while
the vortex model does qualitatively exhibit a strong suppression of the
topological susceptibility as temperature is raised, on a quantitative
level, this suppression is significantly less abrupt than the one found
in $SU(3)$ lattice Yang-Mills measurements. This stands in contrast to
the $SU(2)$ random vortex world-surface model investigated previously
\cite{m2}, which is quantitatively compatible with the corresponding
$SU(2)$ Yang-Mills theory even in the deconfined phase. At this point,
not sufficient information is available to pinpoint the source of the
discrepancy found in the $SU(3)$ case. Possible causes are the following:

The discrepancy may be an artefact of the hypercubic description, which,
as discussed in section~\ref{ambisec}, engenders a significant systematic
uncertainty in the determination of the topological charge. Indeed, this
uncertainty is considerably larger in the $SU(3)$ case than in the
$SU(2)$ case, as evidenced by the change of vortex density in the course
of the vortex world-surface deformations applied, cf.~section~\ref{ambisec},
in order to remove ambiguities in the surfaces. Essentially, the constraint
implied by the hypercubic description, namely, that only six discrete
space-time planes are available in which world-surfaces can extend, is
considerably more stringent in the $SU(3)$ model. Since vortex flux is
fragmented into smaller units in the $SU(3)$ case, already to begin with,
a higher proportion of lattice elementary squares is occupied by vortex
flux; this makes it correspondingly more difficult to find deformations
of the surfaces, within the restricted set of space-time planes, suited
to remove surface ambiguitites. Further to this issue, it should also be
recognized that the topologies of the surfaces qualitatively differ for
$SU(3)$ and $SU(2)$; only the former permits vortex branching. As a result,
comparing the behavior of the world-surface ambiguities in the two cases
is not straightforward; it is entirely possible that the rough
phenomenological prescription applied in section~\ref{ressec} to estimate
the systematic uncertainty, namely, scaling by the appropriate power of
$\rho_{def} /\rho_{orig} $, is subject to sizeable corrections which may
behave very differently for $SU(3)$ and $SU(2)$. It should again be
emphasized that the aforementioned effects are not generic to the
random vortex world-surface concept, but are introduced by the specific
hypercubic construction of the world-surfaces. More comments follow below
on how this limitation may be circumvented.

To the extent that artefacts of the hypercubic description do not account
for the discrepancies compared to Yang-Mills theory observed in the deconfined
phase, several physical causes are possible. For one, as one raises the
temperature past the deconfining phase transition and further into the
deconfined phase, one rapidly approaches the ultraviolet cutoff of the
model. This may appreciably distort the results (again, in a fashion
which may differ considerably for the $SU(3)$ and $SU(2)$ cases due to
the qualitatively different topology of the world-surfaces). The good
agreement with lattice Yang-Mills theory found for $SU(2)$ at high
temperatures could, in this context, be coincidental to an extent.

Furthermore, the pure vortex world-surface curvature action employed in
the model investigated here may not completely capture all relevant
dynamics. Indeed, on general grounds, one expects a shift in the
dynamical characteristics of center vortices as the number of colors $N$
in the underlying Yang-Mills theory is raised \cite{jeffstef}. The
necessity for such a shift, moreover, was observed explicitly in the
$SU(4)$ random vortex world-surface model investigated in \cite{su4}.
There, already the confinement properties of the corresponding
Yang-Mills theory could only be reproduced by introducing new terms
into the effective vortex action beyond a pure world-surface curvature
term. In the present $SU(3)$ case, the confinement properties obtained
using a pure world-surface curvature action still do not differ significantly
from $SU(3)$ Yang-Mills theory, cf.~\cite{su3conf,su3bary,su3freee}.
Possibly, the topological susceptibility investigated in the present
work is more sensitive to the details of the vortex dynamics, and the
discrepancy compared to $SU(3)$ Yang-Mills theory observed at high
temperatures may thus signal the need for an adjustment of the effective
vortex dynamics already for $SU(3)$.

One possibility which comes to mind in this respect is the inclusion
of explicit action terms for thickened monopole world-lines residing
on the vortex world-surfaces. Variational estimates for the energies
of various monopole-type objects have been given in \cite{cw1}; the
objects referred to as ``monopoles'' in the present work correspond to
the ``fully non-Abelian nexi'' of \cite{cw1}. It is argued there
that $SU(3)$ monopoles have considerably higher energy than $SU(2)$
monopoles. This mechanism would therefore be consistent with the phenomena
observed in the random vortex world-surface model in more than one
way: On the one hand, it seems plausible that the deconfined phase is
less entropy-dominated than the confined phase, since the former is,
effectively, dimensionally reduced compared to the latter. Thus, one
would expect the energetics of, say, monopoles to be more relevant in the
deconfined phase, where, indeed, the discrepancies compared to Yang-Mills
theory are observed. Furthermore, if $SU(3)$ monopoles imply a much larger
action penalty than $SU(2)$ monopoles, this would simultaneously explain
why the discrepancies are only observed in the $SU(3)$ model and not the
$SU(2)$ model. Finally, the fact that monopoles are intimately tied to the
topological properties of vortex world-surfaces (by encoding their
non-orientedness) would seem consistent with the discrepancies becoming
apparent specifically in the topological susceptibility.

It would certainly be useful to be able to discriminate between the
different possible sources of the deviation observed at high temperatures
between the $SU(3)$ random vortex world-surface model studied here and
$SU(3)$ Yang-Mills theory. In particular, it would be interesting to
distinguish whether this deviation is an artefact of the hypercubic
construction of the world-surfaces or whether it has a more physical
origin. To this end, an alternative formulation in terms of random
triangulations offers itself, with a minimal area of the triangles
serving as the ultraviolet cutoff. By providing a continuum of directions
into which the vortex surfaces can extend, the most pernicious
world-surface ambiguities discussed in section~\ref{ambisec} would be
avoided from the outset; intersections between surfaces would generically
occur at isolated space-time points. This would obviate the need to 
transfer configurations to a finer scale and perform deformations,
thus avoiding the associated renormalization uncertainties. The
topological charge can be well defined at the original scale in such
a formulation. The drawback of a construction in terms of triangulations
is that book-keeping of vortex locations is considerably more unwieldy,
raising questions concerning, e.g., appropriate rules for dynamically
disconnecting and fusing world-surfaces. Given that the deconfining
phase transition is a percolation transition, such processes play a crucial
role in the vortex picture.

A complementary line of investigation which would mitigate some of the
artefacts engendered by the hypercubic construction, without abandoning
it altogether, lies in evaluating the index of the Dirac operator in the
model vortex ensemble. This can be achieved without transferring vortex
configurations to finer lattices, thus again avoiding the associated
renormalization uncertainties. On the other hand, it must be emphasized
that, in light of the discussion in section~\ref{quantsec}, the
generalization of the index theorem to generic vortex configurations
is not obvious. Since the configurations are defined on manifolds with
a complicated, dynamic topology, corrections to the index theorems valid
on simple manifolds such as spheres or tori must be expected; this
already manifests itself in the fractionalization of global topological
charge found for the model vortex configurations studied here. It would
certainly be interesting to study the correlation between the topological
charge and the Dirac operator index in the vortex ensemble. A construction
of the Dirac operator in the context of the $SU(2)$ vortex world-surface
model has been given in \cite{m3}, albeit discarding the exact zero-mode
sector (which is irrelevant for the chiral condensate discussed there).
A suitable generalization permitting the study of the spectral flow of that
type of operator is needed in order to access the Dirac operator index.

Finally, on a more formal level, it would be interesting to understand
the extent to which the specific half-integer quantization of global
topological charge observed for the model vortex configurations investigated
in this work is determined by the hypercubic description. Surfaces composed
of elementary squares in hypercubic fashion only constitute a particular
subset of continuum surfaces in four dimensions; further fractionalization
of topological charge for more general surfaces is conceivable. This
more general case remains to be explored.

\section*{Acknowledgments}
This work was supported by the U.S.~DOE under grant number DE-FG02-96ER40965.

\end{document}